\documentclass[a4paper]{article}
\usepackage{graphicx}
\usepackage{amsmath,amssymb}
\newcommand\beq{\begin{equation}}
\newcommand\eeq{\end{equation}}
\newcommand\bea{\begin{eqnarray}}
\newcommand\eea{\end{eqnarray}}
\begin{document}

\vspace{-2.0cm}
\bigskip
\begin{flushright} 
Sokendai-HKTK/061202
\end{flushright} 

\centerline{\Large \bf On Loop States in Loop Quantum Gravity}
\vskip 1.0 true cm

\begin{center}
{\bf N. D. Hari Dass}\footnote{hari@soken.ac.jp}, {\bf Manu Mathur} 
\footnote{manu@boson.bose.res.in} 
\vskip 0.6 true cm

${}^{1}$ Hayama Center for Advanced Studies\\
 Hayama, Kanagawa 240-0193, Japan.

\vskip 0.5 true cm

${}^{2}$ S. N. Bose National Centre for Basic Sciences \\
JD Block, Sector III, Salt Lake City, Calcutta 700091, India

\end{center} 
\bigskip

\vskip 0.5 true cm

\centerline{\bf Abstract}

We explicitly construct and characterize all possible independent 
loop states in 3+1 dimensional loop quantum gravity by regulating it on a 3-d 
regular lattice in the Hamiltonian formalism. These loop states, 
characterized by the (dual) angular momentum quantum numbers, describe 
SU(2) rigid rotators on the links of the lattice. The loop states are 
constructed using the Schwinger bosons which are harmonic oscillators 
in the fundamental (spin half) representation of SU(2). Using generalized 
Wigner Eckart theorem, we compute the matrix elements of the volume operator 
in the loop basis. Some simple loop eigenstates of the volume operator are 
explicitly constructed.

\newpage 

\section{Introduction} 

Spin networks were first constructed by Roger Penrose in 1970 to 
give a quantum mechanical interpretation of geometry of space \cite{penrose}. 
Spin networks consist of a minimal subset of  loop states in 
SU(2) gauge theories which solve the Mandelstam constraints and thus provide a complete 
and linearly independent basis in the gauge invariant Hilbert space. The 
loop states  or gauge invariant states, in general, have been extensively studied in 
the context of gauge theories, both in the continuum \cite{mans} as well as on lattice 
\cite{lgt,lgt1,manu2} and  topological field theories \cite{tft}. In particular,  
in \cite{lgt} a complete labelling of the gauge invariant basis states of d+1 
dimensional SU(N) lattice gauge theories was given in terms of the local observables.   
The SU(2) spin networks, in the context of loop quantum gravity (LQG) \cite{rev,rovelli}, 
were first constructed in \cite{lqg} by Rovelli and Smolin. They represent discrete 
quantum states of the 
gravitational field on a 3-d manifold and lead to interesting predictions of discrete 
length, area and volume \cite{rovelli,asht,rovsmol,thiemann}. 
In the present work, we regulate Ashtekar's Hamiltonian formulation 
of gravity on a 3-d regular lattice \cite{rs,rs1,loll} and 
present an alternative approach to construct the SU(2) spin networks 
in terms of the Schwinger bosons \cite{schwinger} which are harmonic 
oscillators in the fundamental representation of SU(2). The 
Schwinger boson creation operators create spin $1/2$ fluxes and are 
the most natural objects to build the spin networks. We further use the 
resulting spin networks or the loop basis  to analyze the spectrum of the 
volume operator. In particular, we explicitly construct some simple loop 
states which are eigenstates of the volume operator of LQG on lattice.  

\noindent In general, for the case of gauge group G, spin networks are 
defined as graphs $\Gamma(\gamma,j,{\cal I})$ where: 
\begin{itemize} 
\item $\gamma$ is a graph with finite set of edges e and a finite set of 
vertices v. 
\item each edge, attached with an irreducible representation of G, 
represents a parallel propagator or holonomy U in that representation. 
\item to each vertex v, we attach an intertwiner ${\cal I}$ to get   
gauge invariant operators at that vertex. 
\end{itemize} 
In the context of LQG, if we restrict ourselves 
to the real connections or holomorphic representation of the Hilbert space 
\cite{rovelli}, the gauge group is SU(2). 
Therefore,  each edge is characterized by spin j. Further, as any 
representation j of SU(2) can be obtained by taking symmetric 
direct product of $2j$ fundamental (spin half) representation,
we also represent the edges as symmetric combination of $2j$ ropes with each 
rope representing the holonomy in the fundamental representation. 
These ropes form closed loops through the intertwiners at the vertices, thus relating 
the spin network graphs to the corresponding loop states.    
The ropes on the edges and the intertwiners at the vertices 
are the basic building blocks of the spin network. 
We find the Schwinger boson approach to spin network very natural as the ropes 
in the fundamental representation are created by the Schwinger boson creation 
operators  (see section 2). Further, it leads to 
the following technical simplifications: a) as Schwinger boson creation 
operators commute amongst themselves there is no need to symmetrize them to get 
higher angular momentum states, b)the  Schwinger 
bosons enable us to construct $k(k-1)/2$ manifestly SU(2) invariant 
intertwining operators at a k valent vertex (see section 2). Thus, 
we avoid the use of the Clebsch-Gordon coefficients in the standard 
spin network construction through the use of holonomies. Therefore, 
we completely bypass the problem of rapid proliferation of 
Clebsch-Gordan coefficients with increasing angular momenta at the 
spin network vertices with given valencies (see section 2.1). 
This proliferation is implicitly contained in the construction 
of spin networks \cite{thiemann,pietri} when they are built out 
of holonomies in the fundamental representation. 

\noindent Infact, using the representation theory of SU(2) group, 
Brunnemann and Thiemann \cite{thiemann} have completely 
characterized the spin network states at a N valent node by 
$(2N-3)$ angular momentum quantum numbers to study the 
spectrum of the volume operator at a vertex. The present paper 
compliments the work of Brunnemann and Thiemann as a) it provides 
a way to explicitly construct these spin network basis states in 
terms of Schwinger bosons, b) it also provides a way to interpret the 
$(2N-3)$ angular momentum quantum numbers in terms of the quantum 
numbers characterizing the corresponding loop states. The relationships 
between group theoretical angular momentum quantum numbers and 
the quantum numbers representing the corresponding loop states are made 
explicit in this work (see section 2.2). These relationships (see (\ref{conslink})), 
in turn, enable us to study the spectrum of volume operator in terms of entire  
loop states. We illustrate these points by explicitly constructing some simple 
loop eigenvectors of the volume operators involving 2 plaquettes on the 
lattice. We also derive the matrix elements of the volume operators 
using generalized Wigner Eckart theorem in the loop basis.  
These matrix elements have been already calculated in \cite{thiemann}. However, 
our method is geometrical and simpler as it exploits the tensor nature of the 
operators through the Wigner Eckart theorem to get the matrix elements 
directly.  

The plan of the paper is as follows. In section 2, we briefly review 
the kinematical variables, their algebras and the constraints in 
lattice LQG \cite{rs,rs1,loll} in the Hamiltonian formulation. This section is for the sake of 
completeness and also  to emphasize the close 
connections between the kinematical variables in LQG and in the Hamiltonian 
lattice gauge theories. More details can be found in \cite{rs,rs1,kogut}. 
In this section, we also discuss the loop states or equivalently spin networks 
in terms of Schwinger bosons.  This discussion is based on the  work done by one of 
the authors \cite{manu2} in the context of loop states in lattice gauge theories. 
In section 3, we investigate the volume operator in lattice LQG. Our volume operator 
on 3-d regular lattice involves only the 3 forward angular momenta. This choice 
coincides in form with the definition of the volume operator of Thiemann 
\cite{thiemann} for a 4-valent 
vertex. However, our vertices on the lattice sites can be up to 6-valent. 
Thus even for the most general states in the Hilbert space of lattice LQG we can
compute the action of the volume operator.
In section (3.1) we compute the matrix elements of the volume operator using generalized 
Wigner Eckart theorem. In section (3.2) we explicitly construct some 
of it's simple eigenvectors in terms of the Schwinger boson operators. 
We also exhibit their loop structure.
  
\section{Hamiltonian lattice gravity} 

In the Ashtekar formulation of Quantum Gravity \cite{rev,rovelli} the kinematical variables are Yang-Mills
gauge field and the associated electric field. The difference from usual 
Yang-Mills theories lies in the constraint structure. In addition to the Hamiltonian and Gauss's 
law constraints 
one has to impose the additional diffeomorphism constraints. Here too one needs to regularise 
the theory. Either the regularisation scheme should manifestly maintain the symmetry and
constraint structures of the theory, or there should be a mechanism to recover these in
the continuum limit. As far as gauge theories are concerned, lattice regularisation is
a non-perturbative regularisation that manifestly preserves gauge invariance, at least as far
vector(as opposed to chiral) theories are concerned. Thus we can expect at least the non-Abelian
gauge invariance of the Ashtekar formalism to be explicitly maintained. At this stage one 
can not say much about diffeomorphism invariance.

In the Ashtekar formulation there is a fiducial space-time on which the gauge fields and their
canonical momenta live. This is not the physical space-time. So one possible way of regularising
the Ashtekar formulation is to adopt a lattice discretisation of the spatial part of this fiducial
space-time.

\begin{figure}[t]
\begin{center}
\includegraphics[width=0.5\textwidth,height=0.12\textwidth]
{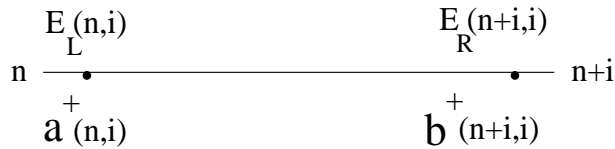}
\end{center}
\vspace{-5mm}
\caption{The assignment of the left and right electric fields 
on a link $(n,i)$. The corresponding Schwinger bosons are also 
 shown by the two $\bullet$ on the link.}   
\label{FIG11}
\end{figure}

Renteln and Smolin \cite{rs,rs1} were the first to put the Ashtekar's Hamiltonian 
formulation of gravity on lattice. We briefly review the nature of the 
kinematical variables involved in this formulation keeping in view the similarities 
with SU(2) Hamiltonian LGT \cite{kogut}. 
%Infact, they  are prcisely 
%the variables involved in pure SU(2) Hamiltonian lattice gauge theory \cite{kogut}.  
We will use a regular cubic lattice as a regulator and denote the lattice sites by 
{\bf n,m} and the 3 directions by {\bf i,j,k} =1,2,3. The SU(2) color index will be denoted by 
a,b,c =1,2,3 and the SU(2) fundamental indices by the Greek alphabets 
$\alpha,\beta,\gamma =1,2$. One associates SU(2) link operators (or holonomies) 
$U_{\alpha\beta}({\bf n,i})$ with the link {\bf (n,i)}. Infact, $U_{\alpha\beta}(n,i)$ can be 
thought of describing the orientation of a SU(2) rigid rotator from body fixed frame 
to space fixed frame. We further introduce the angular momentum operators $E_{L}^{a}(n,i)$ 
and $E_{R}^{a}(n,i)$  which produce the left (body fixed frame) and the right (space 
fixed frame) rotations respectively. Therefore, the canonical commutation relations 
involved are locally that of a rigid body \cite{landau,kogut,rs} and are given by: 
%% NOTE: we should either adopt Loll's notation are give a translation in an Appendix%%%%%%%
\bea 
\left[U_{\alpha\beta}({\it l}),U_{\gamma\delta}({\it l}^{\prime})\right]  =  0, && 
\left[U_{\alpha\beta}({\it l}),U^{\dagger}_{\gamma\delta}({\it l}^{\prime})\right]  =  0, 
\nonumber \\
\label{cr} 
\left[E_{L}^{a}({\it l}),E_{L}^b({\it l}^{\prime})\right]  =  i 
\delta_{{\it l},{\it l}^{\prime}} 
\epsilon^{abc}E_{L}({\it l})  
,  && 
\left[E_{R}^{a}({\it l}),E_{R}^b(m,j)\right]  =   i 
\delta_{{\it l},{\it l}^{\prime}} 
\epsilon^{abc}E_{R}({\it l}), \\ 
\left[E_{L}^{a}({\it l}),U_{\alpha\beta}({\it l}^{\prime})\right]  =  
i \delta_{{\it l},{\it l}^{\prime}} 
\left(\frac{\sigma^a}{2}\right)_{\alpha\gamma}
U_{\gamma\beta}({\it l}), &&  
\left[E_{R}^{a}({\it l}),U_{\alpha\beta}({\it l}^{\prime})\right]  =  i 
\delta_{{\it l},{\it l}^{\prime}} 
U_{\alpha\gamma}({\it l}) 
\left(\frac{\sigma^a}{2}\right)_{\gamma\beta}. \nonumber 
\eea
In (\ref{cr}), ${\it l} = (n,i)$ and ${\it l}^{\prime} = (m,j)$ and  
$\delta_{{\it l},{\it l}^{\prime}} = \delta_{n,m}\delta_{i,j}$.  
The commutation relations (\ref{cr}) clearly show that the 
$E_{L}({\it l})$ and $E_{R}({\it l})$ generate left (at site n) 
and right (at site $(n+i)$) gauge rotations on the holonomy 
$U(n,i)$. Therefore, it is convenient to attach them with the left 
and the right ends of the link $({\it l})$ respectively as shown in 
Figure (\ref{FIG11}). We write them 
as as $E_{L}({\it l}) = E_{L}(n,i) 
= \sum_{a=1}^{3}\sigma^{a}E^{a}_{L}(n,i)$ and $E_{R}({\it l}) 
= E_{R}(n+i,i) = \sum_{a=1}^{3} \sigma^{a} E^{a}_{R}(n+i,i)$. 
The  SU(2) gauge transformations are: 
\bea 
E_{L}(n,i) & \rightarrow & \Lambda(n)E_{L}(n,i)\Lambda^{-1}(n), \nonumber \\
\label{gt} 
E_{R}(n+i,i) & \rightarrow & \Lambda(n+i)E_{R}(n+i,i)\Lambda^{-1}(n+i)   \\ 
U(n,i) & \rightarrow & \Lambda(n)U(n,i)\Lambda^{-1}(n+i) \nonumber 
\eea
Note that the commutation relations (\ref{cr}) are invariant under (\ref{gt}) 
and $E_{L}(n,i)$ and $E_{R}(n+i,i)$ on the link $(n,i)$ gauge transform  by 
$\Lambda(n)$ and $\Lambda(n+i)$ respectively. 
The transformations (\ref{gt})  along with (\ref{cr}) imply that the generator 
of the gauge transformation at lattice site n is: 
\bea 
{\cal C}^{a}(n) \equiv \sum_{i=1}^{3} E^{a}_{L}(n,i) + \sum_{i=1}^{3} E^{a}_{R}(n-i,i) 
\label{gl} 
\eea 
Further, as in the case of rigid body, the body fixed and 
space fixed components of the angular momentum commute 
amongst themselves and their magnitudes are equal, i.e:  
\bea 
\sum_{a=1}^{3}E_{L}^{a}(n,i)E_{L}^{a}(n,i)=\sum_{a=1}^{3}E_{R}^{a}(n+i,i)E_{R}^{a}(n+i,i), ~~ 
\forall ~~(n,i).  
\label{cons} 
\eea 
In the current literature no attention seems to have been paid to these important
constraints. These constraints are crucial at promoting discussion at a single site 
to the entire lattice.
In the present work, we focus  on the Gauss law constraint\footnote{The Hamiltonian 
and diffeomorphism constraints are given in \cite{rev,rovelli,rs,rs1}.} 
(\ref{gl}), ${\cal C}^{a}=0$,  which implement SU(2) gauge 
invariance leading to the 
spin networks.  To solve these constraints \cite{manu2}, we use the Schwinger 
boson representation of SU(2) Lie algebra \cite{schwinger} and write: 
\begin{eqnarray}
\label{sb}
E_{L}^{a}(n,i)  \equiv  a^{\dagger}(n,i)\frac{\sigma^{a}}{2}a(n,i);~~
E_{R}^{a}(n+i,i) & \equiv &  b^{\dagger}(n+i,i)\frac{\sigma^{a}}{2}b(n+i,i).
\end{eqnarray}
on every link (n,i). In (\ref{sb}), $\sigma^{a} (a=1,2,3)$ are the Pauli matrices. 
The SU(2) gauge transformations (\ref{gt}) immediately imply 
that the Schwinger bosons $a_{\alpha}(n,i)$ and $b_{\alpha}(n+i,i)$ transform as 
fundamental representation of SU(2): 
\bea
a_{\alpha}(n,i)\rightarrow\Lambda(n)_{\alpha\beta}
a_{\beta}(n,i); ~~b_{\alpha}(n,i)\rightarrow\Lambda(n)_{\alpha\beta}
b_{\beta}(n,i).
\label{gt3}
\eea
The LQG (like SU(2) lattice gauge theory) in terms of the Schwinger bosons also has a local U(1) 
gauge invariance \cite{manu2} on the lattice links: 
\bea
a_{\alpha}(n,i)\rightarrow exp(i\phi(n,i)) a_{\alpha}(n,i); ~~b_{\alpha}(n+i,i)\rightarrow 
exp(- i\phi(n,i)) b_{\alpha}(n+i,i).
\label{agt}
\eea
The generator of this abelian gauge transformation is: 
\bea
{\cal C}(n,i) = a^{\dagger}(n,i)\cdot a(n,i) - b^{\dagger}(n+i,i)\cdot b(n+i,i)
\label{u1gl}
\eea
where $a^{\dagger}.a \equiv a^{\dagger}_{1}a_{1}+a^{\dagger}_{2}a_{2}.$
Thus, working with the fundamental spin half representation, we have $SU(2) \otimes U(1)$ 
gauge invariance. The corresponding Gauss law constraints are: ${\cal C}^{a} = {\cal C} =0$. 

This underlying $U(1)$ gauge invariance is important in our formulation. 
Without this the construction of the gauge invariant states is incomplete. 
With the choice of Hamiltonian regularisation on a regular 3-d lattice, all the vertices of 
any graph/loop state are at the most of valence 6 and this is a considerable simplification 
over the general situation. For the same reason, the lattice regulator also simplifies the 
the Mandelstam constraints considerably. Further, the present Schwinger boson approach  
to gauge theories also enables us to cast the Mandelstam constraints in local form  
and solve them explicitly \cite{manu2} in terms of the loop states (\ref{fr}). 
We briefly discuss this issue in section (2.2).  A detailed analysis is given in 
\cite{manu3}. 

\noindent In this work we specifically look at the volume operator of Ashtekar and Lewandowski
in loop quantum gravity \cite{asht,thiemann} on the lattice and study its 
matrix elements  and properties of it's spectrum in the loop basis. Even within the 
lattice regularisation there are several choices  for the volume operator that have been 
currently considered in the literature. The one originally considered by Loll \cite{loll} 
is where only the three 'forward' electric fields at each lattice site are involved in the volume
operator at each site. We choose this definition for our present analysis. Technically this corresponds
to the 4-valent vertex volume operator studied by Thiemann et. al. in \cite{thiemann}. If we extend 
the range of ${\bf i}$ from 3 to 6
where for $ i \le 3$ we have $E_L$ at a site and for $i > 3$ we have $E_R$ at that site, the
other choices of the volume operator are
\bea  
V(\Gamma) = \sum_{v \in \Gamma} \sqrt{\frac{1}{3!} \vert{\sum_{i,j,k=1}^{6} 
\epsilon(\hat{i}\hat{j}\hat{k})
\epsilon_{abc}E^{a}[n,i]E^{b}[n,j]E^{c}[n,k]}\vert}    
\label{vol1} 
\eea
In (\ref{vol1}), $\epsilon(\hat{i}\hat{j}\hat{k}) = sign(\hat{i}\times\hat{j}.\hat{k})$. Loll \cite{loll2} has also
used another definition which is 
\bea  
V(\Gamma) = \sum_{v \in \Gamma} \sqrt{\frac{1}{3!} \vert{\sum_{i,j,k=1}^{6} 
|\epsilon(\hat{i}\hat{j}\hat{k})|
\epsilon_{abc}E^{a}[n,i]E^{b}[n,j]E^{c}[n,k]}\vert}    
\label{vol2} 
\eea
which amounts to replacing the 'forward' electric fields at each site by a symmetric sum of 'forward'
and 'backward' electric fields.

All these definitions agree with each other in the naive continuum limit. According to Loll the
difference between the symmetrised and original definitions is only an additive constant for 4-valent
vertices. But for the 6-valent vertices considered here the matrix elements of the
symmetrised expression of Loll, though they still involve only the equivalents of
Thiemann's four valent operators, are more complicated but still calculable in closed form. We shall not
carry out that here but postpone it to a later paper.

Giesel and Thiemann \cite{giesel}  have discussed some consistency conditions that volume operators have to satisfy
but they do not seem to have compared the above alternatives from lattice discretisation.

\subsection{The rigid rotator states on the spin network edges} 

An edge or a link $({\it l})$ of the spin network with quantum number 
$j$ describes the rigid rotator with angular momentum $j$. 
The rigid body constraints (\ref{cons}) or equivalently ${\cal C} =0$ 
in terms of Schwinger bosons, imply that on any link the total occupation number 
of a- type oscillators is same as that of b- type oscillators:  
The states of the rigid rotator \cite{landau} on the link $(n,i)$ 
are characterized by the eigenvalues of $E_{L}(n,i).E_{L}(n,i) (= 
E_{R}(n+,i).E_{R}(n+i,i)), E^{a=3}_{L}(n,i), E^{a=3}_{R}(n+i,i)$. 
We denote the common eigenstate of these 3 operators by 
$|j(n,i),m(n,i),\tilde{m}(n+i,i)>$. 
These states are: 
\bea
|j,m,\tilde{m}> \equiv \frac{(a^{\dagger}_1)^{j+m}(a^{\dagger}_2)^{j-m}
(b^{\dagger}_1)^{j+\tilde{m}}(b^{\dagger}_2)^{j-\tilde{m}}}{\sqrt{(j+m)!
(j-m)!(j+\tilde{m})!(j-\tilde{m})!}} | {}^{0~~0}_{0~~0} \rangle.
\label{hs}
\eea
In (\ref{hs}), $a^{\dagger}_{\alpha} \equiv a^{\dagger}_{\alpha}(n,i)$ and 
$b^{\dagger}_{\alpha} \equiv b^{\dagger}_{\alpha}(n+i,i)$ are defined on 
the left and right end of the link $(n,i)$ respectively. 
Note that a {\it rope} of the spin network represents the first excited state 
$(j= 1/2)$ of the SU(2) rigid rotator. The states $|j,m,\tilde{m}>$ can also be created
by holonomies instead of Schwinger bosons:  
\bea 
|j,m,\bar{m}\rangle = \sum_{S_{2j}} U_{\alpha_{1}\beta_1}
U_{\alpha_{2}\beta_{2}}...U_{\alpha_{2j} \beta_{2j} } |0 \rangle 
\label{lcrr}
\eea
where  the magnetic quantum numbers $\alpha_{i}, \beta_{i} =\pm\frac{1}{2}$ should add upto 
$m$ and $\bar{m}$ respectively, $S_{2j}$ is the
permutation group with $(2j)!$ elements which act on the indices $(\alpha_1,
\alpha_2,....\alpha_{2j})$ to symmetrize and produce higher angular momentum 
states. We note that the Schwinger boson construction (\ref{hs}) is much simpler as it 
does not require any permutation group or equivalently Clebsch-Gordan coefficients whose 
number will increase with increasing angular momentum quantum number j on each link. 
  
\subsection{The intertwining operators at the spin network vertices} 

\begin{figure}[t]
\begin{center}
\includegraphics[width=0.3\textwidth,height=0.28\textwidth]
{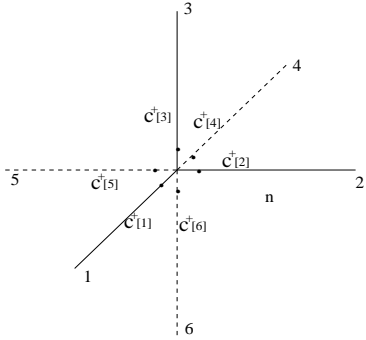}
\end{center}
\vspace{-5mm}
\caption{A graphical representation of the Schwinger boson operators 
$c^{\dagger}[n,i]$, i=1,2,...,6 which are associated with site n. They 
all transform as SU(2) doublets at site n and are shown by $\bullet$ 
on the corresponding axis.}   
\label{FIG22}
\end{figure}
A link with angular momentum $j$ can also be thought of as being represented
by $2j$ independent 'ropes'. Then at a given vertex we have ropes associated
with each of the links. On the other hand a typical loop state consists of a
set of closed loops. Thus to connect the angular momentum basis to loop states
we must have rules on how the ropes on different links at a site are to be 
'tied together' or 'intertwined'; these are given by quantum numbers which we 
call `linking numbers' or  `intertwining numbers'. 
As we show below these quantum numbers arise naturally in the 
Schwinger boson formalism (see \cite{manu2} for more details).   

To define SU(2) gauge invariant intertwining operators at a lattice site n, 
we notice that the site n is associated with 6 Schwinger 
bosons: $a^{\dagger}(n,i)$ and $b^{\dagger}(n,i)$; i=1,2,3,
all gauge transforming as SU(2) doublets.  Therefore, 
we can label them collectively as $c^{\dagger}[n,i]$ with i=1,2,..,6. More explicitly, 
$c^{\dagger}[n,1] =a^{\dagger}(n,1), c^{\dagger}[n,2] =a^{\dagger}(n,2), 
c^{\dagger}[n,3] =a^{\dagger}(n,3), c^{\dagger}[n,4] =b^{\dagger}(n,1), 
c^{\dagger}[n,5] =b^{\dagger}(n,2), c^{\dagger}[n,6] =b^{\dagger}(n,3)$ 
as shown in Figure (\ref{FIG22}). We also relabel the corresponding angular 
momentum operators associated with lattice site n by $J[n,i] 
\equiv c^{\dagger}[n,i]\frac{\sigma^{a}}{2}c[n,i]$ 
with i =1,2,..,6. Note that $J[n,i] = E_{L}(n,i)$  and $J[n,3+i] = E_{R}(n-i,i)$ 
( now for $i=1,2,3$) and the Gauss law (\ref{gl}) now takes the simple form: 
\bea 
{\cal C}^{a}(n) \equiv \sum_{i=1}^{6} J^{a}[n,i] =0 
\label{gls}
\eea 
and simply states that the sum over the 6 angular momenta at any vertex is zero. 
The eigenvalues of $J[n,i]\cdot J[n,i]$ will be denoted by $j(n,i)(j(n,i)+1)$.   
Consider the intertwining of a single rope on the ith link with
a single rope on the jth link. Such a state must be created from the
oscillator vacuum by the action of an operator involving $c_\alpha^\dagger(n,i)c_\beta^\dagger(n,j)$.
Any such operator must be $SU(2)$-invariant.
We are thus led to the following intertwining operator at site $n$:
\bea 
L_{ij}(n) \equiv \epsilon_{\alpha\beta} 
c^{\dagger}_{\alpha}[n,i]c^{\dagger}_{\beta}[n,j] 
= c^{\dagger}[n,i].\tilde{c}^{\dagger}[n,j], i,j =1,2..,6.  
\label{int} 
\eea
In (\ref{int}), $\epsilon_{\alpha\beta}$ is completely antisymmetric tensor 
($\epsilon_{11}=\epsilon_{22}= 0, \epsilon_{12}=-\epsilon_{21}= 1$) and 
$\tilde{c}^{\dagger}_{\alpha} \equiv \epsilon_{\alpha\beta}
{c}^{\dagger}_{\beta}$. Note that: 
\bea 
[{\cal C}^{a}(n),L_{ij}(n)] = 0 
\label{igl}
\eea
and $L_{ij}(n) = -L_{ji}(n)$, $L_{ii}=0$ implying  self intertwining is 
not allowed. Therefore, we can choose the 15 operators $L_{ij}(n)$ with 
$i<j$ to span the space of SU(2) gauge invariant intertwining operators at site n. 
We conclude that any SU(2) gauge invariant state at site n must be of the form: 
\bea 
|\vec{l}(n)> \equiv \left\vert \begin{array}{cccccc}
l_{12} & l_{13}  & l_{14}&l_{15}& l_{16}     \\
 & l_{23} & l_{24} & l_{25} &l_{26} \\
 &  &  l_{34} & l_{35} &l_{36} \\
 &  &  & l_{45} & l_{46}  \\
 &  &  &  & l_{56}   \\
\end{array} \right \rangle =  \prod_{{}^{{i},{j}=1}_{{j} >
{i}}}^{6} \left(L_{ij}(n)\right)^{l_{ij}(n)}
|0>,  ~~ l_{ij}(n) \in {\cal Z_{+}}.
\label{giv2}
\eea
Note that the 
states in (\ref{giv2}) are the eigenstates of the 
angular momentum operators $J[n,i].J[n,i]|_{i=1,2..,6}$ with eigenvalues 
$j[n,i] = j_i(n)(j_i(n)+1)$ 
such that: 
\bea 
2j_{i}(n) = \sum_{k=1}^{6} l_{ik}(n) 
\label{part} 
\eea
This relation follows on counting the number of creation operators of the relevant type in
eqn(\ref{giv2}). It has the following very simple geometrical interpretation:
if we draw $2j_{i}(n)$ lines on the link $[n,i]$ then the 
quantum numbers $l_{ij}$ are the linking numbers connecting the $i^{th}$ and $j^{th}$ 
types of flux lines. Consequently they satisfy $l_{ij} = l_{ji}$. 
To get orthonormal loop states, we choose the complete set of 
commuting operators \cite{lgt,thiemann,manu2} to be 
$J[n,i].J[n,i]|_{i=1,2,..,6}, \left(J[n,1]+J[n,2]\right)^{2}, 
\left(J[n,1] +J[n,2]+J[n,3]\right)^{2},...., J_{total}$ $= \big(J[n,1]+J[n,2]+..$
$+J[n,6]\big)^{2}, 
J_{total}^{a=3}$. We  denote their  common eigenvector by $|j_i(n)|_{i=1,2,..,6}, j_{12}(n),$ 
$j_{123}(n),j_{1234}(n),j_{12345},$  $j_{total},m_{total} \rangle$.  Noting that SU(2) Gauss law 
implies $j_{total}=m_{total} = 0$ and $j_{12345}=j_{6}$. Thus  the common eigenvectors 
are characterized by 9 quantum numbers at every lattice site. This characterization 
of the orthonormal basis by angular momentum operators has been done both in LGT 
\cite{lqg} and LQG \cite{thiemann}. In \cite{manu2} they were  explicitly constructed
and are given by
\begin{eqnarray}
\label{fr}
|j_i(n)|_{i=1,2,..,6},j_{12}(n),j_{123}(n),j_{1234}(n)\rangle  
 =  N(j) \sum_{\{l(n)\}}\hspace{-0.05cm}
{}^{{}^\prime} \prod_{{}^{i,j}_{i < j}} 
\frac{1}{l_{ij}(n)!} 
\big({L}_{ij}(n)\big)^{l_{ij}(n)} |0> 
\end{eqnarray}
The prime over the summation $\{l\}$ 
means that the 15 linking numbers $l_{ij}$ are summed over such that 
the linking numbers $l_{ij}(n)$ satisfy (\ref{part}) along with the 
following constraints: 
\bea 
l_{12} & = & j_1 + j_2 - j_{12} \nonumber   \\
l_{13} + l_{23} & = & j_{12}+j_{3}-j_{123} \nonumber \\ 
l_{14}+l_{24}+l_{34} & = & j_{123}+j_4-j_{1234} \nonumber \\ 
l_{15}+l_{25}+l_{35}+l_{45}&  =&  j_{1234}+j_5-j_{6} 
\label{conslink} 
\eea 
In (\ref{fr}), 
$N(j)  =  N(j_1,j_2,j_{12})N(j_{12},j_{3},j_{123})
N(j_{123},j_4,j_{1234})$$ N(j_{1234},j_5,j_{12345} (=j_{6})) 
N(j_{12345}(=j_6),j_6,0)$ where 
$N(a,b,c)  =  
\Big[\frac{(2c+1)}{(a+b+c+1)!}\Big]^{\frac{1}{2}}
\Big[{(-a+b+c)!(a-b+c)!(a+b-c)!}\Big]^{\frac{1}{2}}$. The constraints (\ref{conslink}) 
are easy to understand: Given $2j_1$ oscillators in direction $(n,1)$ and $2j_2$ oscillators 
in direction $(n,2)$, we need to intertwine (or antisymmetrize) $l_{12}$ oscillators from 
each of these two directions to get a state with angular momentum $j_{12}$. Therefore, 
$(2j_1-l_{12}) + (2j_2-l_{12}) = 2j_{12}$, which is the first equation in (\ref{conslink}). 
Similarly the other equations in (\ref{conslink}) can be obtained.    
Thus we see the ease with which the intertwiners, so crucial to the construction of 
loop states both in lattice gauge theories and loop quantum gravity, emerges from the 
Schwinger boson construction.

We wish to point out here that the above orthonormal basis at every site in conjunction with
the intertwiners satisfying the relations above completely eliminates the notorious problem
of the Mandelstam constraints.
In particular, the primed summation in eqn(\ref{fr}) picks a specific combination of
loop states at each vertex that forms an orthonormal set. This is at the heart of the 
resolution of the Mandelstam constraints. 

\section{The volume operator} 

In this section we study the volume operator in the basis (\ref{fr}) and construct some 
simple loop eigenstates of the volume operator. In \cite{asht,thiemann} the 
volume operator associated with a spin network 
$\Gamma(\gamma,j,v)$ is defined as: 
\bea  
V(\Gamma) = \sum_{v \in \Gamma} \sqrt{\frac{1}{3!} \vert{\sum_{\hat{i},\hat{j},\hat{k}=1}^{6} 
\epsilon(\hat{i}\hat{j}\hat{k})
\epsilon_{abc}E^{a}[n,\hat{i}]E^{b}[n,\hat{j}]E^{c}[n,\hat{k}]}\vert}    
\label{vol} 
\eea
In (\ref{vol}), $\epsilon(\hat{i}\hat{j}\hat{k}) = sign(\hat{i}\times\hat{j}.\hat{k})$. 
Our choice of the volume operator on lattice corresponds to choosing only the 
forward angular momenta at each lattice site \cite{loll}, i.e: 
\bea  
V(\Gamma) &  = &  \sum_{n} \sqrt{\frac{1}{3!} \vert{\sum_{i,j,k=1}^{3}\sum_{a,b,c=1}^{3}  
\epsilon_{ijk} \epsilon_{abc}J^{a}[n,i]J^{b}[n,j]J^{c}[n,k]}\vert} \nonumber \\
& = & \sum_{n} \sqrt{\vert \sum_{a,b,c=1}^{3} 
\epsilon_{abc}J^{a}[n,1]J^{b}[n,2]J^{c}[n,3]\vert} 
\equiv \sum_{n} \sqrt{|Q(n)|}
\label{voll}
\eea
In what follows we will study the spectrum of $Q(n) 
\equiv \epsilon_{abc} J^{a}[n,1] J^{b}[n,2] J^{c}[n,3]$. 
The local operator Q(n) has been extensively studied in the case of 4-valent vertex 
by Brunnemann and Thiemann \cite{thiemann} by writing it as $Q(n) = \frac{i}{4} \hat{q}_{123}(n)$ 
where $\hat{q}_{123}(n) \equiv 
\left[\left(J[n,1]+J[n,2]\right)^{2},\left(J[n,2]+J[n,3]\right)^{2}\right]$. 
Note that we can also write the local operator $Q(n)$ in terms of the 
basic SU(2) gauge invariant intertwining operators (\ref{int}) at site n as $Q(n) = 
\frac{i}{4}\sqrt{\left[L^{\dagger}_{12}(n)L_{12}(n),L^{\dagger}_{23}(n)L_{23}(n)\right]}$.   

\subsection{The matrix elements} 

To calculate the matrix elements of $Q(n)$, we first define the 
angular momentum operator in terms of their tensor or spherical 
components: 
\bea 
J^{(1)}_{\pm1} \equiv \mp \frac{1}{\sqrt{2}}(J^{x}\pm i J^{y}), ~~~~ 
J^{(1)}_{0}\equiv J^{z}. 
\label{tc}
\eea
The irreducible components of the direct product of two tensors  
$A^{(j_1)}_{m_1}$ and $ B^{(j_2)}_{m_2}$ of rank $j_1$ and $j_2$ 
respectively with $m_1 (m_2)$ varying from $-j_1 (-j_2)$ to $j_1 (j_2)$,   
are defined as \cite{varsha}: 
\bea 
\left[A^{(j_1)} \times B^{(j_2)}\right]^{(j_{12})}_{m_{12}} \equiv 
\sum_{m_1,m_2} C^{j_{12}m_{12}}_{j_1,m_1;j_2,m_2} A^{(j_1)}_{m_1} B^{(j_2)}_{m_2}. 
\label{dp2t} 
\eea
In (\ref{dp2t}), $C^{j_{12}m_{12}}_{j_1,m_1;j_2,m_2}$ are the Clebsch- Gordon 
coefficients \cite{varsha} and $m_{12}$ vary from $-j_{12}$ to $+j_{12}$. The 
scalar product of two tensors of the same rank is denoted by: 
\bea 
A^{(j)} \cdot B^{(j)} \equiv \left[A^{(j_1=j)} \times B^{(j_2=j)}\right]^{(0)}_0 
=  \sum_{m_1,m_2} C^{j_{12}=0 m_{12}=0}_{j,m_1;j,m_2} A^{(j)}_{m_1} B^{(j)}_{m_2}
= \sum_{m=-j}^{+j} \frac{(-1)^{(j-m)}}{\sqrt{((j))}} A^{(j)}_m B^{(j)}_{-m}.  
\label{sp}
\eea
In (\ref{sp}), $((j)) = (2j+1)$ represents the multiplicity factor.  
Using the definitions (\ref{dp2t}) and (\ref{sp}), we write: $Q(n) \equiv 
\epsilon_{abc} J^{a}[n,1] J^{b} [n,2]
J^{c}[n,3] = +i\sqrt{2((1))} (J^{(1)}[n,1] \times J^{(1)}[n,2])^{(1)}
\cdot(J^{(1)}[n,3])^{(1)}$. Now the matrix elements of $Q(n)$ can be directly written 
down using generalized Wigner Eckart theorem \cite{varsha}:
\bea 
&& <j_1,j_2,j_{12},j_3;j_{123},m_{123}|\Big[(J^{(1)}[n,1] \times J^{(1)}[n,2])^{(1)}\cdot J^{(1)}[n,3]
\Big]| \bar{j}_1,\bar{j}_2,\bar{j}_{12},\bar{j}_3;\bar{j}_{123},\bar{m}_{123}> \nonumber \\  
&& = (-1)^{j_{123}-m_{123}}
{\left(\begin{array}{cccc}
{j}_{123} & 0 & \bar{j}_{123}   \\
-{m}_{123} & 0 & \bar{m}_{123}\\
\end{array} \right)}
[((0))((j_{123}))((\bar{j}_{123}))]^{\frac{1}{2}} 
{\left \{\begin{array}{cccc}
{j}_{12} & \bar{j}_{12} & 1   \\
{j}_{3} & \bar{j}_{3} & 1\\
{j}_{123} & \bar{j}_{123} & 0 \\  
\end{array} \right \}} \nonumber \\
&&\left(j_1,j_2,j_{12}||\left(J^{(1)}[n,1]\times J^{(1)}[n,2]\right)^{(1)}||
\bar{j}_1,\bar{j}_2,\bar{j}_{12}\right) 
\left(j_3||J^{(1)}[n,3]||\bar{j}_3\right) 
\label{gwe1}
\eea
In (\ref{gwe1}), $ {\left(\begin{array}{cccc} j_1 & j_2 & j_3  \\ {m}_{1} & m_2 & m_3\\ \end{array} 
\right)}$ and ${\left \{\begin{array}{cccc}
{j}_{1} & {j}_{2} & j_{12}   \\
{j}_{3} & {j}_{4} & j_{34}\\
{j}_{13} & {j}_{24} & j \\  
\end{array} \right \}}$ are the Clebsch Gordan and ${9 j}$ symbols respectively. 
%The $((j)) \equiv 2j+1$ represents the multiplicity  and 
The reduced matrix elements of the operator J are represented by $(j||J||\bar{j})$. 
Similarly, the reduced matrix element $(j_1,j_2,j_{12}||(J^{(1)}[n,1] \times 
J^{(1)}[n,2])^{(1)}||\bar{j}_1,\bar{j}_2,\bar{j}_{12})$ are given by: 
\bea 
\left(j_1,j_2,j_{12}||\left(J^{(1)}[n,1] 
\times J^{(1)}[n,2]\right)^{(1)}||\bar{j}_1,\bar{j}_2,\bar{j}_{12}\right)
 & = &  \big[((1))((j_{12}))((\bar{j}_{12}))\big]^{\frac{1}{2}} 
{\left \{\begin{array}{cccc}
{j}_{1} & \bar{j}_{1} & 1   \\
{j}_{2} & \bar{j}_{2} & 1\\
{j}_{12} & \bar{j}_{12} & 1 \\  
\end{array} \right \}}  ~~~~~~\nonumber \\
&& \left(j_1||J^{(1)}[n,1]||\bar{j}_1\right)~~ \left(j_2||J^{(1)}[n,2]||\bar{j}_2\right)
\eea
Now using the values: 
\bea 
(j||J||\bar{j}) & = & \delta_{j,\bar{j}}\left[j(j+1)(2j+1)]\right]^{\frac{1}{2}} 
\equiv \delta_{j,\bar{j}}x(j) \nonumber \\ \nonumber \\
{\left(\begin{array}{cccc}
{j}_{123} & 0 & \bar{j}_{123}   \\
-{m}_{123} & 0 & \bar{m}_{123}\\
\end{array} \right)} 
& = & (-1)^{j_{123}-m_{123}}  
\frac{1}{\sqrt{((j_{123}))}} ~\delta_{j_{123},\bar{j}_{123}}
\delta_{m_{123},\bar{m}_{123}} \nonumber  \\ \nonumber \\
{\left \{\begin{array}{cccc}
{j}_{1} & {j}_{1} & 1   \\
{j}_{2} & {j}_{2} & 1\\
{j}_{12} & \bar{j}_{12} & 1 \\  
\end{array} \right \}} 
& = & (-1)^{j_1+j_2+j_{12}} \frac{\left[j_{12}(j_{12}+1)-\bar{j}_{12}(\bar{j}_{12}+1)\right]}
{2\sqrt{2((1))}x(j_1)}  
{\left \{\begin{array}{cccc}
{j}_{12} & \bar{j}_{12} & 1   \\
{j}_{2} & {j}_{2} & j_1\\
\end{array} \right \}} \nonumber \\ \nonumber \\
{\left \{\begin{array}{cccc}
{j}_{12} & \bar{j}_{12} & 1   \\
{j}_{3} & \bar{j}_{3} & 1\\
{j}_{123} & \bar{j}_{123} & 0 \\  
\end{array} \right \}}  
&=& \frac{(-1)^{j_{12}+j_3+j_{123}}}{\sqrt{((1))((j_{123}))}} {\left \{\begin{array}{cccc}
{j}_{12} & \bar{j}_{12} & 1   \\
{j}_{3} & {j}_{3} & j_{123}\\
\end{array} \right \}} 
\eea 
and putting them in (\ref{gwe1}), we get:  
\bea 
\label{eff} 
&& <\bar{j}_{i}|_{i=1,.,6},\bar{j}_{12},\bar{j}_{123},\bar{j}_{1234}|Q(n)
|j_{i}|_{i=1,.,6},j_{12},j_{123},j_{1234}> 
 =  \frac{i}{2} (-1)^{(j_1+j_2+j_{12}+\bar{j}_{12}+j_3+j_{123})} 
~~~~~~~\nonumber \\ 
&&
\left(\prod_{\hat{i}=1}^{6} \delta_{\bar{j}_{\hat{i}},j_{\hat{i}}}\right) 
\delta_{\bar{j}_{123},j_{123}}  \delta_{\bar{j}_{1234},j_{1234}}  
{x(j_2,j_3)} y(j_{12},\bar{j}_{12}) 
 {\left\{ \begin{array}{cccc}
\bar{j}_{12} & {j}_{12} & 1  \\
{j}_{2} & j_{2} & j_1\\
\end{array} \right \}}
{\left\{ \begin{array}{cccc}
\bar{j}_{12} & {j}_{12} & 1  \\
{j}_{3} & j_{3} & j_{123}\\
\end{array} \right \}}  
\nonumber \\
&& \equiv 
\left(\prod_{\hat{i}=1}^{6} \delta_{\bar{j}_{\hat{i}},j_{\hat{i}}}\right) 
\delta_{\bar{j}_{123},j_{123}}  \delta_{\bar{j}_{1234},j_{1234}} <\bar{j}_{12}|Q(n)|j_{12}>  
\eea
where, 
\bea 
x(j_2,j_3)  & \equiv &  x(j_2)x(j_3) = \sqrt{j_2(j_2+1)(2j_2+1)j_3(j_3+1)(2j_3+1)} \nonumber \\
y(j_{12},\bar{j}_{12})  & \equiv &  \sqrt{((j_{12}))((\bar{j}_{12}))}. 
\left[j_{12}(j_{12}+1) - \bar{j}_{12}(\bar{j}_{12}+1)\right]. 
\label{eqns11} 
\eea
We note that the matrix elements (\ref{eff}) are antisymmetric and imaginary as the operator 
$Q(n)$ is hermitian. 
Putting the explicit values of the 6-j symbols we get: 
\bea 
<j_{12}+1|Q(n)||j_{12}>  & = &  \frac{+i}{4\sqrt{((j_{12}))((j_{12}+1))}} 
\Big[{(j_1+j_{12}+j_{2}+2)(j_{123}+j_{12}+j_{3}+2)} \nonumber \\
&& {(j_1+j_{12}-j_2+1) (j_{123} +j_{12}-j_{3}+1)
(j_1-j_{12}+j_2)(j_{123}-j_{12}+j_{3})} \nonumber  \\
&& {(-j_1+j_{12}+j_2+1) (-j_{123}+j_{12}+j_{3}+1)}\Big]^{\frac{1}{2}} \nonumber \\
& = & - <j_{12}|Q(n)||j_{12}+1>.    
\label{xxyy}
\eea
The last result follows because the matrix elements of Q(n) are antisymmetric and 
purely imaginary. As mentioned in the introduction, the matrix elements in 
(\ref{eff}) and (\ref{xxyy}) are already obtained in \cite{thiemann} by 
writing $Q(n) \equiv \frac{i}{4} \left[\left(J[n,1]+J[n,2]\right)^{2},
\left(J[n,2]+J[n,3]\right)^{2}\right]$. This computation required a change 
of basis from $|j_1,j_2,j_{12},j_3,j_{123},m_{123}>$ to 
$|j_{1},j_2,j_3,j_{23},j_{123},m_{123}>$ and  use of 
Elliot-Biedenharn identity. The analysis  
of this section, based on the Wigner-Eckart theorem, is direct and also 
quite general. 

\begin{figure}[t]
\begin{center}
\includegraphics[width=0.98\textwidth,height=0.3\textwidth]
{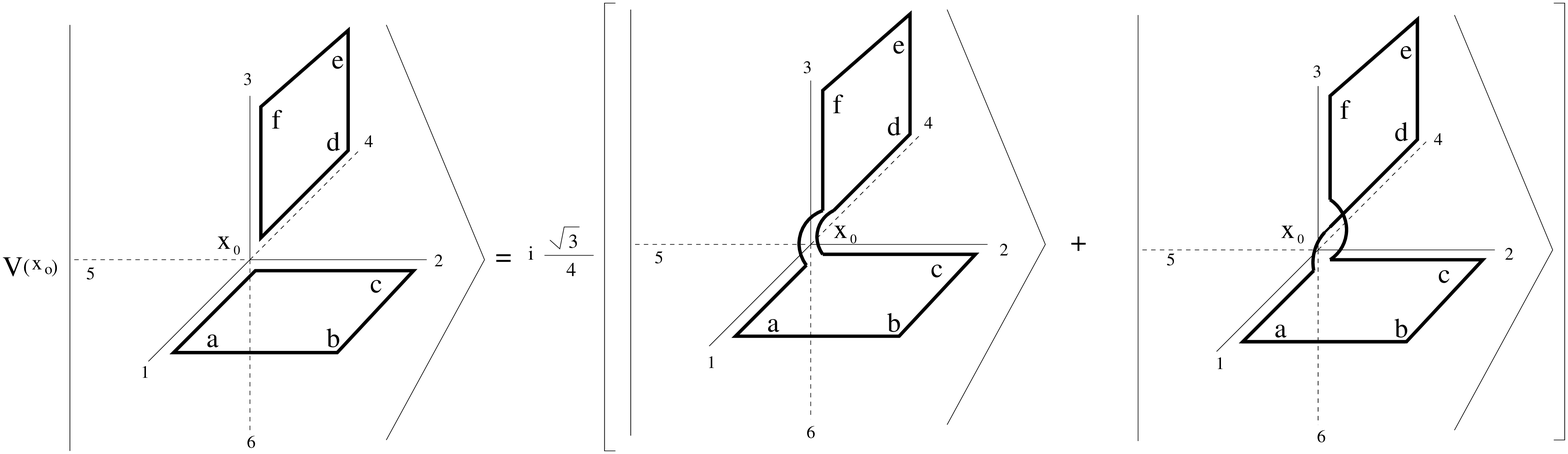}
\end{center}
\vspace{-5mm}
\caption{Graphical representation of the action of the operator Q(n) (\ref{xxcc}) on the 
simple loop state $|j_{12}=0>_{{x_{0}}} \otimes |abcdef>$ in (\ref{ls}). The loop state 
on the right hand side represents: $|j_{12}=1>_{{x_{0}}} \otimes |abcdef> \equiv 
|1/2,1/2,1/2,1/2,0,0,1,1/2,0>_{x_{0}} \otimes |abcdef>$. Note that the operator Q(n) 
changes the linking number: $l_{12} \rightarrow l_{12} \pm 1$ at $x_{0}$. The other 
linking numbers are changed in accordance with eqn(\ref{conslink}).} 
\label{FIG33}
\end{figure}

\subsection{An example} 

In this section, we use an example to illustrate the simplest but non trivial action of the 
volume operator on an entire loop state. We also work out the corresponding loop eigenstates 
of the volume operator.  As all the planar loops are trivially annihilated by the volume operator 
and therefore have zero volume, we need to consider loops spread over all the  3 forward 
directions. This implies that the simplest graph  with non-zero volume will involve 
2 plaquettes in two different planes. Further,  we associate j= 1/2 with all the links 
of the two plaquettes centered at a lattice site $x_{0}$ as shown in the Figure (\ref{FIG33}). 
The corresponding loop state $|LS>$ is direct product of the states 
$|j_{1},j_{2},j_{3},j_{4},j_{5},j_{6}, j_{12}, j_{123}, j_{1234}>$ in (\ref{fr}) at the vertices 
($x_{0},a,b,c,d,e,f$). 
\bea 
|LS> & = & 
|1/2,1/2,1/2,1/2,0,0,0,1/2,0>_{x_{0}} 
\otimes
|0,1/2,0,1/2,0,0,1/2,1/2,0>_{a} 
\nonumber \\ 
&& \otimes
|0,0,0,1/2,1/2,0,0,0,1/2>_{b} 
\otimes
|1/2,0,0,0,1/2,0,1/2,1/2,1/2>_{c} 
\nonumber \\ 
&& \otimes
|1/2,0,1/2,0,0,0,1/2,0,0>_{d} 
\otimes
|1/2,0,0,0,0,1/2,1/2,1/2,1/2>_{e} 
\nonumber \\ 
&& \otimes
|0,0,0,1/2,0,1/2,0,0,1/2>_{f} \equiv 
|j_{12}=0>_{{x_{0}}} \otimes |abcdef>  
\label{ls} 
\eea
Using (\ref{xxyy}) we find that: 
\bea 
\label{xxcc}
Q(x_{0}) |1/2,1/2,1/2,1/2,0,0,0,1/2,0>_{x_{0}} & = & i~\frac{\sqrt{3}}{4}~
 |1/2,1/2,1/2,1/2,0,0,1,1/2,0>_{x_{0}}  \nonumber \\
Q(x_{0}) |1/2,1/2,1/2,1/2,0,0,1,1/2,0>_{x_{0}} & = & -i\frac{\sqrt{3}}{4}~
 |1/2,1/2,1/2,1/2,0,0,0,1/2,0>_{x_{0}}   \\ \nonumber \\
Q(n) |j_1,j_2,j_3,j_4,j_5,j_6,j_{12},j_{123},j_{1234}>_{n}  & = & 0; ~~~ n \neq x_{0}.
\nonumber  
\eea
Note that the states at a,b,c,d,e and f are all annihilated by the volume operator
and  $Q = - \frac{\sqrt{3}}{4} \sigma_{2}$ in the loop  space of this example where 
$\sigma_{2}$ is the Pauli matrix. 
It is also instructive to use this example to clarify the meaning of $l_{ij}$. First
consider the state occurring on the lhs of the first equation of eqn(\ref{xxcc}); it
is easy to check on using eqn(\ref{part}) and eqn(\ref{conslink}) that for this
state $l_{12}=l_{34}=1$ and all other $l_{ij}$'s are zero. Likewise, for the state
occurring on the rhs of this equation there are two possibilities: $l_{13}=l_{24}=1$
all the rest being zero, and $l_{14}=l_{23}=1$ and all the rest being zero. It also
follows from eqn(\ref{fr}) that both these states occur with equal weight and they 
have non-zero overlap. Thus the top of eqn(\ref{xxcc}) at $x_{0}$ can be graphically 
represented as shown in Figure 3.

Thus the two simplest loop eigenstates of the volume operator are: 
\bea 
|+\frac{{\sqrt{3}}}{4}>&=&\frac{1}{\sqrt{2}} \big(~|j_{12}=0>_{x_{0}}\otimes~|abcdef> + i~
|j_{12}=1>_{x_{0}}\otimes~|abcdef>\big)   
\nonumber \\
|-\frac{\sqrt{3}}{4}>&=&\frac{1}{\sqrt{2}} \big(~|j_{12}=0>_{x_{0}}\otimes~|abcdef> - i~
|j_{12}=1>_{x_{0}}\otimes~|abcdef>\big)   
\label{esvo} 
\eea
with the property:~~$ V|\pm \frac{\sqrt{3}}{4}> = \Big[\sum_{n}~\sqrt{|Q(n)|}\Big] |\pm 
\frac{\sqrt{3}}{4}> = \sqrt{\frac{\sqrt{3}}{4}} |\pm \frac{\sqrt{3}}{4}>$.  These degenerate 
loop eigenstates can be explicitly constructed in terms of the Schwinger bosons 
using (\ref{fr}). 

\section{Summary and Discussion} 

The main objective of the present work is to emphasize the utility of 
spin half Schwinger bosons in explicitly constructing and characterizing 
all the $SU(2) \otimes U(1)$ gauge invariant loop states in loop quantum gravity. 
In particular, the reformulation in terms of Schwinger boson enables us to interpret 
the SU(2) gauge invariant states of \cite{lgt,lgt1,thiemann} in the (dual) angular momentum 
representation in terms of the geometrical loops. This approach is also technically useful as 
it is completely in terms of gauge invariant intertwining quantum numbers defined in  section 
2.2 and also  bypasses the problem of rapid proliferation of Clebsch Gordan coefficients  in 
constructing the spin networks using holonomies.  Further, we have  given 
a simple derivation of the matrix elements of the volume operator in LQG 
by using generalized Wigner Eckart theorem. The simplest loop eigenstates 
of the volume operator are explicitly constructed.   

We note that in the present approach the  SU(2) Gauss law constraints 
(\ref{gls}) and their independent solutions (\ref{fr}) are defined at the 
lattice {\it sites}. Next are the the U(1) Gauss law (\ref{u1gl}) constraints 
which act on the {\it links}. In this logical sequence, the next set of constraints 
are the diffeomorphism and the Hamiltonian constraints which  involve the holonomies and 
electric field operators  over the {\it plaquettes} \cite{rs,rs1} of the lattice. 
Therefore, our next objective is  to study the diffeomorphism and the Hamiltonian 
constraints in the loop basis (\ref{fr})  over the entire lattice. 
It will also be interesting to cast the Hamilitonian and the diffeomorphism constraints 
in terms of Schwinger bosons or equivalently in terms of the invariant intertwiners 
discussed in section (2.2). These issues are currently under investigation.      
 
{\bf Acknowledgment:} One of the authors (M.M) would like to thank Professor Binayak 
Dutta Roy for useful discussions.

\end{document}